\documentclass[aps,prl,twocolumn,showkeys,showpacs,superscriptaddress,groupedaddress]{revtex4}  
\usepackage{graphicx}  
\usepackage{dcolumn}   
\usepackage{bm}        
\usepackage{amssymb}   
\usepackage{latexsym,bm}
\usepackage[tbtags]{amsmath}

\begin{document}

\widetext


\title{The Spatial Scaling Laws of Compressible Turbulence}

\author{Bohua Sun}

\affiliation{%
Department of Mechanical Engineering\\
Cape Peninsula University of Technology, Cape Town, South Africa\\
sunb@cput.ac.za
}%

\date{\today}

\begin{abstract}
\small
The spatial scaling laws of velocity kinetic energy spectrum for compressible turbulence flow and its density-weighted counterpart have been formulated in terms of wavenumber, dissipation rate and Mach number by using dimensional analysis. We have applied the Barenblatt's incomplete similarity theory to both kinetic and density-weighted energy spectrum and showed that, within the initial subrange, both energy spectrums approach the -5/3 power law of the wavenumber, when the Mach number $M$ tends to be naught, unity and infinity, respectively.
\end{abstract}

\maketitle
\section{Introduction}

Turbulence is considered to be one of the unsolved problems in classical physics \cite{kri,frisch,lee,she,zhou,wang2013,zhang1,chen2015}. There are two kinds of turbulence; one is incompressible and another is compressible. In recent years compressible turbulence has drawn a great deal of attention. Fully developed three-dimensional homogenous incompressible turbulence has been studied by Kolmogorov \cite{k41a,k41b,k41c} who showed its energy spectrum exhibits $k^{-\frac{5}{3}}$ power of wavenumber $k$ in the inertial subrange. However, the basic processes, which occur in compressible turbulence are less understood (Aluie \cite{alu2011,alu2012,alu2013}; Armstrong \textit{et al} \cite{ams}; Cardy, \textit{et al} \cite{cardy}; Chu \textit{et al} \cite{chu}; Federrath, \textit{et al}\cite{fed}; Kritsuk \cite{kri}; Schmidt \cite{schmidt}; Galtier \cite{gal};  Wang \emph{et al} \cite{wang2013,wang2014} ; Sun \cite{sun2015,sun2016b}).

Kovasznay (1963) \cite{kov} points out that the compressible turbulence problem is characterized by the existence of acoustic, vortical and entropy modes when in interaction with each other. Systematic 2nd-approximation of these nonlinear interactions was shown by Chu and Kovasznay (1958) \cite{chu}. Lighthill and Whitham (1955) \cite{light} argue that its energy is continually radiated away in the form of sound waves, which are ultimately converted into heat by the various processes of acoustic attenuation. Therefore, one may visualize the compressibility effects as acting like a source of energy dissipation in addition to that, which is provided by viscosity and thermal conductivity (Moiseev, \textit{et al.} \cite{moiseev}). As the nonlinear effects become prominent, the sound waves in a compressible fluid sharpen to form shock waves, while vortex formation behind the shock waves then produces anisotropic shear turbulence. The investigation showed that the passage of a shock wave also results in smaller parallels to the shock and are compressed in the direction perpendicular to the shock. The shock formation and the shock interaction process lead to another source of energy dissipation in compressible turbulence.

Despite the anisotropy which is caused by the individual shock, Kadomtsev and Petviashvili (1973) \cite{kad} argue that the random orientation of the various shocks leads to the overall isotropy of the turbulent field. Using a Burgers equation type model, Kadomtsev and Petviashvili (1973) \cite{kad} provide a spectrum for kinetic energy, which is shown below
\begin{equation}\label{e1}
E(k) \sim \varepsilon c^{-1}k^{-2}.
\end{equation}
where $c$ is the speed of sound in the fluid, $k$ is wave number and $\varepsilon$ is the dissipation rate.

Moiseev \textit{et al.}(1981) \cite{moiseev} applied group-invariance principles to the Hopf-type functional formulation of the compressible case, and gave the spectrum of the kinetic energy, as shown below
\begin{equation}\label{e2}
E(k)\sim \rho c^{\frac{2}{3\gamma-1}}\varepsilon^{\frac{2\gamma}{3\gamma-1}}k^{-\frac{5\gamma-1}{3\gamma-1}}.
\end{equation}
where $\rho$ is the density of the fluid and $\gamma$ is the ratio of specific heats of the fluid.

Shivamoggi\cite{shiva} asserts that Equation (\ref{e2}) is not completely correct and proposes a revised spectrum as
\begin{equation}\label{e3}
E(k)\sim \rho^{\frac{\gamma-1}{3\gamma-1}} c^{\frac{2}{3\gamma-1}}\varepsilon^{\frac{2\gamma}{3\gamma-1}}k^{-\frac{5\gamma-1}{3\gamma-1}}.
\end{equation}

However, it is easy to verify that the previous scaling laws, namely Equations (\ref{e1},\ref{e2} and \ref{e3}), have a common problem with presentation of the speed of sound $c$ and density $\rho$, which are supposed to be in the dimensionless form such as $\rho_0/\rho$ or $v/c$. Therefore, the scaling laws of the compressible turbulence proposed in the Equations (\ref{e1},\ref{e2} and \ref{e3}) might be not a proper format and should be investigated further.

High resolution numerical simulations for supersonic isothermal turbulence showed the $-\frac{5}{3}$ spectrum of the density-weighted velocity ${\bf v} =\rho^{1/3} \bf{u}$, where $\rho$ is the density and $\bf{u}$ is the velocity (Krituk, \textit{et al.} \cite{kri}; and Schmidt, \textit{et al.} \cite{schmidt}). Aluie \cite{alu2011,alu2012} proved that kinetic energy cascades conservatively in compressible turbulence, provided that the pressure-dilatation co-spectrum decays at a sufficiently rapid rate, which was supported once again by numerical simulations (Wang \textit{et al.} \cite{wang2013}). Recently, for Mach number $M$ near $1$ the numerical simulation from (Wang \textit{et al.} \cite{wang2013}) confirmed that fully developed three-dimensional compressible turbulence density-weighted energy spectrums exhibited Kolmogorov's $-\frac{5}{3}$ power law in the inertial range. Galtier and Banerjee \cite{gal} derived a relation for the scaling of compressible isothermal turbulence, and showed only around the sonic scale, when the local Mach number dropped to unity, and the density-weighted energy spectrum approached the $-\frac{5}{3}$ power law.

Since the numerical results can only be obtained for a specific case, there is no way to predict the general scaling trends on the energy spectrums, alternative ways have to be proposed. From author's point of view, there are three questions that should be answered. The first is how to predict scaling laws for an arbitrary Mach number. The second relates to whether the velocity kinetic energy $E(k,\varepsilon,M)$ can still be used to characterise the cascade of compressible turbulence, and the final question is in what form the scaling laws of the density-weighted energy spectrum $E_\rho(k,\varepsilon, M)$ for an arbitrary Mach number $M$ will be.

Due to the complicated nature of the compressible turbulence, the above questions may not be fully answered by numerical simulations, hence alternative ways should be sought. Although the nature of the compressible turbulence is still not fully understood, we believe that its physics must satisfy the dimensional laws. The results in this article show that the dimensional analysis can certainly capture the overall picture of the compressible turbulence cascade process, and leads to fairly rich information on the phenomena.

One of the most successful applications of dimensional analysis is for turbulence problem was made by Kolmogorov \cite{k41a,k41b,k41c}. For homogenous incompressible turbulence, Kolmogorov introduced the length, time and velocity scales of the smallest eddies of turbulence, whereas the smallest scales are given by the Kolmogorov length $\eta=(\varepsilon^3/\nu)^{1/4}$, the energy-dissipation rate $\varepsilon=2\nu \overline{s_{ij}s_{ij}}$, $s_{ij}=\frac{1}{2}{(u'_{i,j}+u'_{j,i}) }^2$, and $u'_i$ fluctuation of flow velocity. Kolmogorov states that the large-scale turbulence motion is roughly independent of viscosity. The small scale, however, is controlled by viscosity. In the inertial range, turbulence is controlled solely by the dissipation rate $\varepsilon$ and the size of eddy $k$. It is found that the spatial energy spectrum $E(k)$ can be formulated in terms of wave number $k$ and dissipation rate $\varepsilon$ as $E(k)=\varepsilon^{2/3}k^{-5/3}f[{(k\eta)}^{4/3}]=C_K\varepsilon^{2/3}k^{-5/3}$, which is the famous Kolmogorov $-\frac{5}{3}$ law of incompressible turbulence and $f(:)$ is a dimensionless function.

To make the paper self-contained, the paper is organised as follows. After this introduction, in Section 2 the spatial scaling law of the kinetic energy spectrum $E$ of velocity $\bm u$ has been formulated by using dimensional analysis, the scaling laws by using the Barenblatt's incomplete similarity theory has been presented, some special cases of the laws have been obtained. In Section 3 the scaling laws of the density-weighted energy spectrum $E_\rho$ of $\rho^{1/3}\bm u$ has been given by the similar way as in section 2. In Section 4 the discussions on the obtained scaling laws have been proposed. Finally, section 5 concludes the paper.

\section{Kinetic Energy spectrum of compressible turbulence for velocity }

In general, compressible flow deals with fluid density, which varies significantly in response to a change in pressure that is caused by high flow speed. Compressibility effects are typically considered to be significant if the Mach number $M$ of the flow exceeds $0.3$. For incompressible fluid there is no need to consider changes in mass density. However, mass density change is the central concern for compressible flow, which must be taken into account in the formulation.

In order to formulate the compressible turbulence scaling law, Sun \cite{sun2015a} extended Kolmogorov's assumption from incompressible turbulence to a compressible case. The idea of this extension stems from the dimensional analysis of the lift of a wing, in which the lift force $F_L$ is a function of velocity $V$, air density $\rho$, wing cross section area $A$, angle of attack $\alpha$, viscosity $\mu$, and speed of sound $c$, which will provide then the lift $F_L=f(V,A,\rho, \mu, c,\alpha)$. According to dimensional analysis then, the lift is $F_L=\frac{1}{2}\rho V^2 A f(Re,M,\alpha)$. This relation is still valid for any Reynolds $Re$ and Mach number $M$.

\subsection{Dimensional analysis and choice of dimensional variables for compressible turbulence}

Any physical relationship can be expressible in a dimensionless form. The implication of this statement is that all of the fundamental equations of physics, as well as all approximations of these equations and, for that matter, all functional relationships between these variables must be invariant under a dilation of the dimensions of the variables. This is because the variables are subject to measurement by an observer in terms of units that are selected at the arbitrary discretion of the observer. It is clear that a physical event cannot depend on the particular ruler, which is used to measure space, the clock is used for time, and the scale is used to measure mass, or any other standard of measure that might be required, depending on the dimensions that appears in the problem. This principle is the basis for a powerful method of reduction, which is called dimensional analysis, and is useful for the investigation of complicated problems. (Bridgman \cite{bridgman1922}; Sedov \cite{sedov}; Barenblatt \cite{baren}; Cantwell \cite{cantwell}; Sun \cite{sun2016}).

Often, dimensional analysis is conducted without any explicit consideration for the actual equations that may govern a physical phenomenon. Only the variables that affect the problem are considered. in fact, this is little deceiving. Inevitably, the choice of the variables is intimately connected to the phenomenon itself and, therefore, is always connected to, and has implications for the governing equations. The most complex problems in dimensional analysis tend to be filled with ambiguity, particularly regarding the choice of variables that govern the phenomenon in question. The success or failure of dimensional analysis depends entirely on the choice of dimensioned physical variables  that are relevant to the problem. This constitutes the art of dimensional analysis. Applied intelligently with a deep knowledge of the problem, may yield important and profound results. Applied blindly, dimensional analysis can easily lead to nonsense.((Bridgman \cite{bridgman1922}; Sedov \cite{sedov}; Barenblatt \cite{baren}; Cantwell \cite{cantwell}; Sun \cite{sun2016}).

From a physics point of view, dimensional analysis is an universal method, which can, of course, be used for the study of compressible turbulence. The difference between incompressible turbulence and its compressible counterpart is that flow mass density $\rho$ will no longer be a constant, because in the compressible case of, the mass density changes as a result of a high speed that will generate shock waves and some of the interactions mentioned in above. Our believe is that no matter how complex the compressible turbulence, as long as we can capture all the primary variables of the problem, then we can formulate it by way of dimensional analysis.

In 1941 the Russian statistician, A. N. Kolmogorov, published three papers (Kolmogorov \cite{k41a,k41b,k41c}) that provide some of the most important and most-often quoted results of incompressible homogeneous turbulence theory. These results comprise what is now referred to as the K41 theory, and represent a major success of the statistical theories of turbulence. This theory provides a prediction for the energy spectrum of a 3D isotropic homogeneous turbulent flow. Kolmogorov proved that even though the velocity of an isotropic homogeneous turbulent flow fluctuates in an unpredictable fashion, the energy spectrum (how much kinetic energy is present on average at a particular scale), is predictable.

In the initial subrange the K41 theory assumes that the spectrum $E$, at any particular wave number, $k$ depends only on the dissipation rate $\varepsilon$, namely, $E=f(k,\varepsilon)$.

Due to the great success of the Kolmogorov theory, it would be natural to attempt to extend the Kolmogorov idea to compressible turbulence. As we know, the basic difference between incompressible and compressible flow concerns the compressibility of mass-density $\rho$. By taking into account the mass-density $\rho$ or the Mach number $M$, we can extend Kolmogorov's assumption to compressible turbulence as follows:

\emph{In the inertial range the compressible turbulence energy spectrum $E$ is not only controlled by the dissipation
rate $\varepsilon$ and the wave number $k$, but also by the fluid density $\rho$ (or the Mach number $M$).}

In the extended Kolmogorov assumption, there are two sets of variables in the formulation, namely one that includes a mass density $\rho$ and the other that has Mach number $M$.

\emph{The set I}: For the first set there are six variables: $E$ the energy spectrum; $\varepsilon$ dissipation rate; $k$ wave number; $\nu$ kinematic viscosity; $\rho$ the current mass density; and $\rho_0 $ the local "reservoir values" of mass density or stagnation density (Liepmann and Roshko \cite{liep}).

All primary dimensions of Set I are listed in the Table\ref{table1}.

\begin{table}[ht]
\caption{The primary dimensions list of set I}\label{table1}
\footnotesize
\centerline{
\begin{tabular}{ccccccc}
\hline
$E$ & $\nu$ & $\varepsilon$ & $k$ & $\rho$ & $\rho_0$ \\
\hline
$L^3t^{-2}$ & $L^2t^{-1}$ & $L^2t^{-3}$ & $L^{-1}$ & $mL^{-3}$ & $mL^{-3}$ \\
\hline
\end{tabular}}
\end{table}

From Liepmann and Roshko \cite{liep}, the density ration $\frac{\rho_0}{\rho}$ and velocity ration $\frac{v}{c}\equiv M$ has a relationship $\frac{\rho_0}{\rho}= [1+\frac{\gamma-1}{2}(\frac{v}{c})^2]^\frac{1}{\gamma-1}$, where $ \gamma =\frac{c_p}{c_v}$ is the ratio of heat capacities. This means that the two ratios are dependent, we can use either $\frac{\rho_0}{\rho}$ or $\frac{v}{c}$ in the formulation. Then we have the set II of dimensional variables.

\emph{The Set II}: For the 2nd set there are six variables: $E$ the energy spectrum; $\varepsilon$ dissipation rate; $k$ wave number; $\nu$ kinematic viscosity; the mass density and its reservoir values can be replaced by fluid velocity $v$ and the speed of sound $c$.

All primary dimensions of Set II are listed in the Table.\ref{table2}.

\begin{table}[ht]
\caption{The primary dimensions list of set II}\label{table2}
\footnotesize
\centerline{
\begin{tabular}{ccccccc}
\hline
$E$ & $\nu$ & $\varepsilon$ & $k$ & $c$ & $v$ \\
\hline
$L^3t^{-2}$ & $L^2t^{-1}$ & $L^2t^{-3}$ & $L^{-1}$ & $Lt^{-1}$ & $Lt^{-1}$ \\
\hline
\end{tabular}}
\end{table}

\subsection{Scaling laws based on set I of dimensional variables}

According to the Buckingham $\Pi$ theorem (Bridgman \cite{bridgman1922}), the energy spectrum $E$ can be expressed as the function of $(\nu,\varepsilon,k,\rho_0,\rho)$
\begin{equation}\label{e4}
E=f(\nu,\varepsilon,k,\rho_0,\rho).
\end{equation}

Within the six variables, there are three basic units, namely time $t$, mass $m$ and length $L$. We can choose three repeating variables, namely wavenumber $k$, dissipation rate $\varepsilon$, and density $\rho$; the dependent variables is the energy spectrum$E$, the kinematic viscosity $\nu$ and $\rho_0$.

From the Buckingham $\Pi$ theorem of dimensional analysis, we have three dimensionless variables, namely $\Pi_1=E\varepsilon^{-3/2}k^{5/3}$, $\Pi_2=\nu\varepsilon^{-1/3}k^{4/3}$, and $\Pi_3=\rho_0\rho^{-1} $, then we have the scaling law of the energy spectrum $\Pi_1=f(\Pi_2,\Pi_3)$, that is
\begin{equation} \label{e5}
E(k,\varepsilon,\rho)=\varepsilon^{2/3}k^{-5/3}f[(k\eta)^{4/3},\rho_0\rho^{-1}].
\end{equation}

\subsection{Scaling laws based on set II of dimensional variables}

If we use the fluid velocity $v$ and sound speed $c$ instead of the mass density, we have the second set of variables $(E, \nu,\varepsilon,k,c,v)$, as well as another version of Equation(\ref{e4}) as follows
\begin{equation}\label{e6}
E=f(\nu,\varepsilon,k,c,v).
\end{equation}

Hence, we can have a set of corresponding dimensionless $\Pi$ which is $\Pi_1=E\varepsilon^{-3/2}k^{5/3}$, $\Pi_2=\nu\varepsilon^{-1/3}k^{4/3}$, and $\Pi_3=v/c=M $, where $M=v/c$ is the Mach number, and the scaling law of the energy spectrum can be expressed in terms of Mach number as $\Pi_1=f(\Pi_2,\Pi_3)$, so $E(k,M)$
\begin{equation}\label{e7}
E(k,\varepsilon,M)=\varepsilon^{2/3}k^{-5/3}f[(k\eta)^{4/3},M].
\end{equation}

$E(k,\varepsilon,M)$ in Equation(\ref{e7}) is actually equivalent to $E(k,\varepsilon,\rho)$ in Equation (\ref{e5}), because the ratio of density in equation(\ref{e5}) can be expressed in terms of Mach number as $\frac{\rho_0}{\rho}= [1+\frac{\gamma-1}{2}{M}^2]^\frac{1}{\gamma-1}$. Hence, the energy spectrum equation(\ref{e5}) can be rewritten as the function of the Mach number equation(\ref{e7}).

It should be pointed out that the dimensionless function $f(x,y)$ cannot be completely determined by only dimensional analysis, which should be finalized by other ways such as numerical or experiments.

\subsection{Scaling laws based on incomplete similarity}

For incompressible turbulence, we can further simplify Equation(\ref{e7}). According to the Kolmogorov's assumption (1941a, 1941b), in the inertial sub-range the term $k\eta\rightarrow 0$, so for the finite value of the Mach number $M$, the function is $f[(k\eta)^{4/3},M] \rightarrow f(0,M)$. Kolmogorov assumed that in the limits $k \eta \rightarrow 0$, the function $f(0,x)$ simply assumes the constant value of $C_K$. In other words, there is complete similarity with respect to the variables $k \eta \rightarrow0$, and hence we have $E(k)=C_K \varepsilon^{2/3}k^{-5/3}$. This is the famous Kolmogorov $-\frac{5}{3}$ spectrum, one of the cornerstone of turbulence theory. $C_K$ is a universal constant, the Kolmogorov constant, experimentally found to be approximately $1.5$.

However, for the compressible turbulence the existence of the limit of $f(x,M)$ as $x \rightarrow 0$ is a question owing to intermittency -- the fluctuations of the energy dissipation rate about its mean value $\varepsilon$. According to Barenblatt\cite{baren}, the incomplete similarity in the variable $k\eta$ would require the nonexistence of a finite and nonzero limit of $f(x,M)$ as $x \rightarrow 0$.

To obtain more information from equation (\ref{e7}), let us use Barenblatt's incomplete similarity theory\cite{baren} to simplify the function $f(x,y)$. For this purpose, we propose two hypotheses for compressible turbulence, as shown below.

\textit{First hypothesis}: There is  incomplete similarity regarding the energy spectrum $E(k,\varepsilon,M)$ in the parameter $k\eta$ and no similarity in the Mach number $M$.

\textit{Second hypothesis}: The energy spectrum $E(k,\varepsilon,M)$ tends to be a well-defined limit as the viscosity tends to be very small but not zero.

In terms of the first hypothesis, for a large $M$, the function $f(x,y)$ could be assumed to be the power function of its argument $k\eta$, while there is no kind of similarity in $M$, as follows
\begin{equation}\label{e8}
f(k\eta,M)=A(M)[(k\eta)^{4/3}]^{B(M)}.
\end{equation}
Then we arrive at the energy spectrum relation
\begin{equation}\label{e9}
E(k,\varepsilon,M)=\varepsilon^{2/3}k^{-5/3}A(M)[(k\eta)^{4/3}]^{B(M)},
\end{equation}
where $A(M)$ and $B(M)$ are a function of the Mach number, and $B(M)$ is the so-called intermittency exponent.

Using a similar approach proposed by Barenblatt \cite{baren}, the $B(M)$ can be further simplified to a linear form $B(M)= \alpha+\beta \epsilon$, where $\alpha,\ \beta$ should be determined by constants, and the asymptotic parameter $\epsilon$ is a function of the March number vanishing when $M \rightarrow \infty$. Then we have
\begin{equation}\label{e10}
E(k,\varepsilon,M)=\varepsilon^{2/3}k^{-5/3}A[(k\eta)^{4/3}]^{\alpha+\beta \epsilon}.
\end{equation}

Applying the second hypothesis, a well-defined limit of $E(k,\varepsilon,M)$ exists only if $\alpha=0$. Applying the mathematic identity $x^a=e^{a\ln x}$ to equation (\ref{e10})), then we have
\begin{equation}\label{e11}
E(k,\varepsilon,M)=\varepsilon^{2/3}k^{-5/3}A e^{4(\beta \epsilon \ln k\eta)/3}.
\end{equation}
In Equation(\ref{e11}), when viscosity vanishes $k\eta \rightarrow 0$, then $\ln k\eta \rightarrow -\infty $. If $\epsilon$ tended to be zero as $M \rightarrow \infty$ faster than $\frac{1}{\ln{M}}$, then the exponent would tend to be zero and we would return to the case of complete similarity, which should not be the case in terms of compressible turbulence. However, if $\epsilon$ tends to be zero and is slower than $\frac{1}{\ln{M}}$, a well-defined limit of the energy spectrum will not exist, which will violate the second hypothesis. Therefore, the only choice, that is  compatible with the hypotheses must be
\begin{equation} \label{e12}
\epsilon=\frac{1}{\ln{M}}.
\end{equation}

The similar result was firstly obtained by Barenblatt \cite{baren}for the boundary turbulence flow, where the small perturbation parameter is $\epsilon=\frac{1}{\ln{Re}}$, where $Re$ is Reynolds number. Sun (2015a, 2015b) extended this idea to compressible turbulence.

Therefore, the Equation(\ref{e11}) would be in the form of
\begin{equation}\label{e13}
E(k,\varepsilon,M)=\varepsilon^{2/3}k^{-5/3}A[(k\eta)^{4/3}]^{\beta/\ln{M}}.
\end{equation}
Substitute the Kolmogorov length $\eta=(\varepsilon^3/\nu)^{1/4}$ into Equation(\ref{e13}), we have
\begin{equation}\label{e14}
E(k,\varepsilon,M)=\varepsilon^{2/3}k^{-5/3}A [(k(\varepsilon^3/\nu)^{1/4})^{4/3}]^\frac{\beta}{\ln{M}},
\end{equation}
or it can be rewritten as in a compact form as
\begin{equation} \label{e15}
E(k,\varepsilon,M)=C\varepsilon^{(\frac{2}{3}+\frac{\beta}{\ln{M}})}k^{(-\frac{5}{3}+\frac{4\beta}{3\ln{M}})},
\end{equation}
where the coefficient is
\begin{equation}\label{e16}
C=A \nu^{-\frac{\beta}{3\ln{M}}},
\end{equation}
in which $C$ appears to be the inversely proportional to the $\ln M$.

For a large but not infinite Mach number, the Equation (\ref{e15}), together with equation (\ref{e16}), is the Barenblatt-type incomplete scaling law for compressible turbulence. The modified exponents  $C$  represent the intermittency of the process, and the $C$ and $\beta$ must be determined by either numerical or experimental data at a given Mach number.

The formulations, which are presented in this section provide answers to the first and second questions mentioned in introduction section.

\subsection{Scaling laws of some special cases}

Despite the unknown constants of $A$ and $\beta$, the Equations (\ref{e15}) and (\ref{e16}) can still yield fairly rich information on how the kinetic energy varies with the wavenumber and the dissipation rate. Although the Equation(\ref{e15}) was formulated for a large Mach number due to the nature of $\ln M$ in Equation (\ref{e12}), we surprisedly found that it is still valid for quite a wide range of Mach number $M$.

With the modified scaling law in Equation(\ref{e15}), we can also obtain scaling laws for the following four limit cases: (i). incompressible turbulence $M=0$; ii). sonic turbulence $M=1$; iii). supersonic turbulence; and iv. hypersonic turbulence.

i. The Mach number is zero, hence $\ln{M} \rightarrow \infty$, so for any $\beta$ the exponents $\frac{2}{3}+\frac{\beta}{\ln{M}}\rightarrow \frac{2}{3}$ and $-\frac{5}{3}+\frac{4\beta}{3\ln{M}} \rightarrow -\frac{5}{3}$, the Equation(\ref{e14}) can be reduced to
\begin{equation}\label{e17}
E(k,\varepsilon,M)|_{M=0}=E(k,\varepsilon,0)= A \varepsilon^{2/3}k^{-5/3}.
\end{equation}

According to the Kolmogorov $-\frac{5}{3}$ power scaling laws of incompressible turbulence, the constant $A$ is equal to the Kolmogorov universal constant $C_K$, i.e., $A=C_K=1.5$.

ii. For sonic turbulence $M=1$, then $\ln M =\ln 1=0$, so in order to have a well-defined limit energy spectrum, the parameter $\beta$ must be zero, that is $\beta=0$. The equation(\ref{e14}) is reduced to the $-\frac{5}{3}$ power law
\begin{equation}\label{e18}
E(k,\varepsilon,M)|_{M=1}=E(k,\varepsilon,1)=A \varepsilon^{2/3}k^{-5/3},
\end{equation}
in which $A$ must be determined by using other approaches.

iii. In the case of  hypersonic turbulence $M \rightarrow \infty$, $\ln{M} \rightarrow \infty$, then $(\frac{2}{3}+\frac{\beta}{\ln{M}}) \rightarrow  \frac{2}{3}$ and $-\frac{5}{3}+\frac{4\beta}{3\ln{M}} \rightarrow -\frac{5}{3}$, the equation(\ref{e14}) can be reduced to
\begin{equation}
E(k,\varepsilon,M)= A \varepsilon^{2/3}k^{-5/3}.
\end{equation}
in which $A$ must be determined by using other approaches.

iv. In the case of supersonic turbulence, the Mach number $M$ tends to a big given number $M_G$ and $\beta$ is given by $\beta=-\frac{1}{4}{\ln{M_G}}$, then $(\frac{2}{3}+\frac{\beta}{\ln{M_G}})=\frac{5}{12}$ and $B(M_G)=-\frac{5}{3}+\frac{4\beta}{3\ln{M_G}} =-\frac{5}{3}-\frac{1}{3}=-2$, therefore the kinetic energy spectrum becomes
\begin{equation}\label{e19}
E(k,\varepsilon,M)=A \nu^{1/12} \varepsilon^{5/12} k^{-2},
\end{equation}
in which the exponent $2$ is bigger than $\frac{5}{3}$ (the Kolmogorov scaling) indicates this case is significantly steeper than Kolmogorov scaling. The obtained scaling law of $k^{-2}$ in (\ref{e19}) is very close to the scaling law of $k^{-1.95}$ by Kritsuk \emph{et al.} \cite{kri}. Furthermore, our Equation (\ref{e19}) also provide the dissipation ratio scaling law as $\varepsilon^{5/12}$, which has not been reported ever before.

\subsection{Generalization of Kritsuk's scaling laws}

As we indicated before, the dimensional analysis can only give an universal result, however, some constants in the result can only be determined by other methods. In Equation (\ref{e15}), there is two unknown adjustable constants $\beta$ and $A$, which should be determined by other approaches.

We can use the numerical results from the famous work done by Kritsuk \emph{et al.} \cite{kri} to determine the constant $\beta$. The scaling law of $k^{-1.95}$ obtained numerically by Kritsuk \emph{et al.} \cite{kri} at Mach number $M=6$, that is $M_G=6$, then we have a good reason to set the constant as
\begin{equation}\label{111a}
 -\frac{5}{3}+\frac{4\beta}{3\ln{M_G}}  \approx -2, \text{ which is close to -1.95}
\end{equation}
which gives
\begin{equation}\label{111}
  \beta=-\frac{1}{4}(\ln{M_G})|_{M_G=6}=-\frac{1}{4}{\ln{6}}=-0.448.
\end{equation}
With the constant $\beta$, the parameter $B(M)$ is given by
\begin{equation}\label{222}
  B(M)=-\frac{5}{3}-\frac{1}{3}\frac{\ln 6}{\ln{M}}.
\end{equation}
Having the $\beta$, we can determined the constant $A$ with the help of the Kolmogorv universal constant $C_K$, which experimentally was determined to be $C_K=1.5$. The idea to find the $A$ is that the constant $C=A\nu^{\frac{\ln 6}{12\ln M}}$ should be equal to $C_K$ in the limits of $\nu \rightarrow 0$ and $M\rightarrow0$.

Set $y=\nu^{\frac{\ln 6}{12\ln M}}$, then $\ln y= \frac{\ln 6}{12}\frac{\ln M} {\ln\nu}$, its limit is
\begin{equation}
\begin{array}{ccc}
\lim \limits_{\nu\rightarrow 0, M\rightarrow0} \ln y=\frac{\ln 6}{12}\frac{\frac{d\ln M}{dM}} {\frac{d\ln \nu}{d\nu}}\\
=\frac{\ln 6}{12}\frac{1/M}{1/\nu}=\frac{\ln 6}{12}\frac{\nu}{M}=\frac{\ln 6}{12}\approx0.14931289.
\end{array}
\end{equation}
Then we have $y=e^{0.14931289}=1.161$, which lead the constant $A=C_K/y=1.5/1.161=1.2919$.

Hence, we can predict the scaling law of compressible turbulence as follows
\begin{equation} \label{e15b}
E(k,\varepsilon,M)=1.2919\nu^{\frac{\ln 6}{12\ln{M}}}\varepsilon^{(\frac{2}{3}-\frac{1}{4}\frac{\ln 6}{\ln{M}})}k^{(-\frac{5}{3}-\frac{1}{3}\frac{\ln 6}{\ln{M}})}.
\end{equation}
This expression is a generalization based on the work of Kritsuk \emph{et al} \cite{kri}. To appreciate Kritsuk \emph{et al} \cite{kri} great contributions to the compressible turbulence, we propose formula (\ref{e15b}) is called Kritsuk's scaling laws.

The scaling laws (\ref{e15b}) is valid for wide range of the Mach number $M$. It is clear for the incompressible turbulence that the exponent $\frac{\ln 6}{\ln M}\rightarrow 0$ and $\nu^{\frac{\ln 6}{12\ln{M}}}\rightarrow 1.161$ when $M \rightarrow 0, \, \nu\rightarrow 0$, the $E(k,\varepsilon,M)$ should be almost independent from $\nu$ for any fixed $k$ belonging to the inertial range, that is, $E(k,\varepsilon,M) \rightarrow E(k,\varepsilon)$
\begin{equation} \label{e15c}
E(k,\varepsilon)=1.5\varepsilon^{\frac{2}{3}}k^{-\frac{5}{3}}.
\end{equation}
It is not surprise to see that the first scaling law of the expression (\ref{e15c}) is the well known Kolmogorov (K41) scaling law for incompressible turbulence which has been presented in the introduction section.

\begin{figure}[ht]
  \centerline{\includegraphics[scale=0.3]{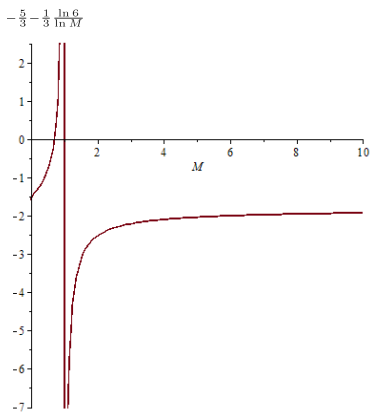}}
\caption{\label{fig01} Curve of $B(M)=-\frac{5}{3}-\frac{1}{3}\frac{\ln 6}{\ln{M}}$.}
\end{figure}

The exponent index $B(M)$ of $k$ is plotted as in Figure \ref{fig01}.
From $B(M)$, it is easy to see the tangent of Log-curve between $\ln E$ and $\ln k$: for $0<M<0.69$, the tangent is negative; for $0.69<M<1$, the tangent is positive and for $1<M<\infty$, the tangent is negative. For a big Mach number, the $B(M)$ tends to be a constant $-\frac{5}{3}$.

\section{Kinetic Energy spectrum of compressible turbulence for the density-weighted velocity}

Due to the density change of compressible turbulence, as a tradition, the numerical simulations usually use the density-weighted velocity ${\bf v} =\rho^{1/3} \bf{u}$ instead of velocity $\bm u$, where $\rho$ is the density and $\bf{u}$ is the velocity.

The corresponding density-weighted energy spectrum $E_\rho$ can be expressed as
\begin{equation}\label{e20}
E_\rho=F(\nu,\varepsilon,k,\rho, M),
\end{equation}

All primary dimensions are listed in the Table \ref{table3}.

\begin{table}[ht]
\caption{The primary dimensions list}\label{table3}
\footnotesize
\centerline{
\begin{tabular}{ccccccc}
\hline
$E_\rho$ & $\nu$ & $\varepsilon$ & $k$ & $\rho$ & $M$ \\
\hline
$m^{1/3}L^2t^{-2}$ & $L^2t^{-1}$ & $L^2t^{-3}$ & $L^{-1}$ & $mL^{-3}$ & $1$ \\
\hline
\end{tabular}}
\end{table}

The problem has six parameters and three primary dimensions $(m,L,t)$, and according to the Buckingham $\Pi$ theorem of dimensional analysis, we will have three $\Pi$ as $\Pi_1=E_\rho \rho^{-1/3}\varepsilon^{-2/3}k^{5/3}$ and $\Pi_2=(k\eta)^{4/3}$, $\Pi_3=M$ and relationship $\Pi_1=f(\Pi_2,M)$, ie.
\begin{equation}\label{e21}
E_\rho(k,\varepsilon,M)=\rho^{1/3}\varepsilon^{2/3}k^{-5/3}F((k\eta)^{4/3},M).
\end{equation}
Equation shows the density-weighted energy spectrum $E_\rho$ has the same power exponents as $E$.

In the same way, the Barenblatt's incomplete similarity theory \cite{baren} can also be applied to Equation (\ref{e21}), which will lead to similar results as before, namely
\begin{equation} \label{e21a}
E_\rho(k,\varepsilon,M)=C_\rho \rho^{1/3}\varepsilon^{(\frac{2}{3}+\frac{\beta_\rho}{\ln{M}})}k^{(-\frac{5}{3}+\frac{4\beta_\rho}{3\ln{M}})},
\end{equation}
where the constants $A_\rho, \, \beta_\rho$ and coefficient $C_\rho=A_\rho \nu^{-\frac{\beta_\rho}{3\ln{M}}}$.

For different scenario of Mach number $M$, we can derive some scaling laws as follows:

i. The Mach number is zero, hence $\ln{M} \rightarrow \infty$, so for any $\beta$ the exponents $\frac{2}{3}+\frac{\beta}{\ln{M}}\rightarrow \frac{2}{3}$ and $-\frac{5}{3}+\frac{4\beta}{3\ln{M}} \rightarrow -\frac{5}{3}$, the Equation(\ref{e14}) can be reduced to
\begin{equation}\label{e17a}
E_\rho(k,\varepsilon,M)|_{M=0}=E_\rho(k,\varepsilon,0)= A_\rho \rho^{1/3} \varepsilon^{2/3}k^{-5/3}.
\end{equation}

ii. For sonic turbulence $M=1$, then $\ln M =\ln 1=0$, so in order to have a well-defined limit energy spectrum, the parameter $\beta$ must be zero, namely $\beta=0$. The equation(\ref{e14}) is reduced to the $-\frac{5}{3}$ power law
\begin{equation}\label{e18a}
E_\rho(k,\varepsilon,M)|_{M=1}=E_\rho(k,\varepsilon,1)=A_\rho \rho^{1/3}\varepsilon^{2/3}k^{-5/3},
\end{equation}
in which $A_\rho$ must be determined by using other approaches.

iii. In the case of supersonic turbulence, the Mach number $M$ tends to a big given number $M_G$, which satisfies $(\frac{2}{3}+\frac{\beta}{\ln{M_G}})=\frac{1}{3}$, then $\beta=-\frac{1}{3}\ln M_G$ and $-\frac{5}{3}+\frac{4\beta}{3\ln{M_G}} =-\frac{5}{3}-\frac{4}{9}=-\frac{19}{9}$, therefore the kinetic energy spectrum becomes
\begin{equation}\label{e19a}
E_\rho(k,\varepsilon,M)=A_\rho \nu^{1/9} \rho^{1/3} \varepsilon^{1/3} k^{-\frac{19}{9}},
\end{equation}
which is same as the scaling law of $k^{-\frac{19}{9}}$ obtained by Galtier and Banerjee \cite{gal} for density-weighted energy spectrum. In the same time, our Equation (\ref{e19a}) also provide the dissipation ratio scaling law as $\varepsilon^{1/3}$, which has not been reported ever before.

iv. In the case of supersonic turbulence, the Mach number $M$ tends to a big given number $M_G$ and $\beta$ is given by $\beta=-\frac{1}{4}{\ln{M_G}}$, then $(\frac{2}{3}+\frac{\beta}{\ln{M_G}})=\frac{5}{12}$ and $B(M_G)=-\frac{5}{3}+\frac{4\beta}{3\ln{M_G}} =-\frac{5}{3}-\frac{1}{3}=-2$, therefore the kinetic energy spectrum becomes
\begin{equation}\label{e19}
E_\rho(k,\varepsilon,M)=A_\rho \rho^{1/3}\nu^{1/12} \varepsilon^{5/12} k^{-2}.
\end{equation}

v. In the case of highly compressible ( $M \rightarrow \infty $), $\frac{2}{3}+\frac{\beta_\rho}{\ln{M}} \rightarrow \frac{2}{3}$ and $-\frac{5}{3}+\frac{4\beta_\rho}{3\ln{M}} \rightarrow -\frac{5}{3}$, Equation (\ref{e21a}) can be simplified to
\begin{equation} \label{e22}
E_\rho(k,\varepsilon,M)=A_\rho \rho^{1/3}\varepsilon^{2/3}k^{-5/3}.
\end{equation}

\begin{figure}[ht]
  \centerline{\includegraphics[scale=0.25]{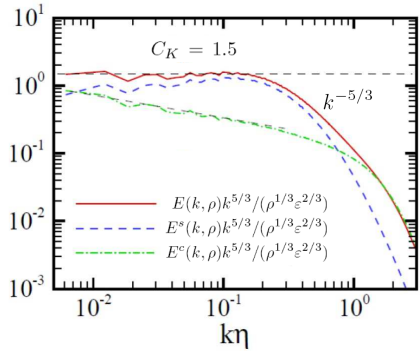}}
\caption{\label{fig1} Density-weighted energy spectrum of compressible turbulence (Wang \cite{wang2014}).}
\end{figure}

The scaling laws demonstrated in Equation (\ref{e22}) was confirmed by numerical simulation by Wang \cite{wang2013,wang2014}as illustrated in the Figure \ref{fig1}. The data of Figure 1 was taken from Wang \cite{wang2014} and reformatted, where $E_\rho=E(k,\rho)$ is total kinetic energy, and $E_\rho^c=E^c(k,\rho)$ and $E_\rho^s=E^s(k,\rho)$ are compressible and solenoidal parts of the $E(k,\rho)$, their relationship is $E(k,\rho)=E^c(k,\rho)+E^s(k,\rho)$.

The $C_K$ was determined by curve fitting as $C_K=1.5$, which is close to the Kolmogorov universal constant. Equation (\ref{e22}) indicates that a density-weighted energy spectrum in $k^{-5/3}$ may still be preserved at small scales ($k\eta\rightarrow 0$) if the density-weighted fluid velocity $\bf{v}=\rho^{1/3}\bf{u}$, is used. This conclusion was firstly mentioned by Wang \emph{et al}\cite{wang2013}.

It might be worth noting that the relationship between the kinetic energy spectrum $E(k,\varepsilon,M)$ and its density-weighted counterpart $E_\rho(k,\varepsilon,M)$ can be established from Equations (\ref{e5}), (\ref{e7}) and (\ref{e21}) as
\begin{equation}\label{e23}
\frac{E_\rho(k,\varepsilon,M)}{E(k,\varepsilon,M)} = \rho^{1/3}\frac{F(k\eta,M)}{f(k\eta,M)}.
\end{equation}
where the functions $F(k\eta,M)$ and $f(k\eta,M)$ are different from each other, and in general their ratio is not a constant.

At small scales ($k\eta\rightarrow 0$) and in a limited case of complete compressibility, $M \rightarrow \infty$ , if the ratio in Equation (\ref{e23}) does exist, becomes
\begin{equation}\label{e24}
\frac{E_\rho(k,\varepsilon,M)}{E(k,\varepsilon,M)} \rightarrow \lambda \rho^{1/3}.
\end{equation}

From numerical simulation at Mach number close to $1$ carried by Wang \cite{wang2014}, the coefficient $\lambda$ can be estimated as $\lambda \simeq 1$.

The relation (\ref{e24}) answerer the question in Section 1: the compressible turbulence can be characterised by the $\rho^{1/3}\bm u$ density-weighted kinetic energy spectrum $E_\rho(k,\varepsilon,M)$, as well as by the $\bm u$ kinetic energy spectrum $E(k,\varepsilon,M)$.

The formulation in this section shows that the density-weighted kinetic energy spectrum can be obtained by dimensional analysis.

\section{Discussions}

This article proposes spatial scaling laws of the kinetic energy spectrum $E(k,\varepsilon,M)$ of compressible turbulence flow and its density-weighted counterpart $E_\rho(k,\varepsilon,M)$ in terms of wavenumber $k$, dissipation rate $\varepsilon$ and Mach number $M$. The study shows that the compressible turbulence kinetic energy spectrum and the density-weighted energy spectrum does not behaves in complete similarity, but rather in incomplete similarity, as in Equations (\ref{e15}) and (\ref{e21}), and both energy spectrums in $k^{-5/3}$ may still be preserved at small scales ($k\eta\rightarrow 0$) for compressible turbulence.

Why can the energy spectrum in $k^{-5/3}$ still be preserved? As we knows, in compressible turbulence, the kinetic energy spectrum can be decomposed into compressible and solenoidal parts: $E=E^c+E^s=c^ck^{-2}+c^sk^{-5/3}=k^{-5/3}(c^ck^{-8/9}+c^s)$, in which the compressible energy spectrum $E^c=c^ck^{-2}$ and the solenoidal energy spectrum $E^s=c^sk^{-5/3}$, $c^c$ and $c^s$ are constants.

There are nonlinear interactions between solenoidal and compressive modes of velocity fluctuations. The numerical simulations showed that the compressive kinetic energy $E^c$ and its solenoidal counterpart $E^s$ follows different cascade scaling laws \cite{wang2013,wang2014}. The compressible kinetic energy $E^c$ cascade follows $k^{-2}$ power laws of the wave number, and the solenoidal $E^s$ follows the $k^{-5/3}$ power laws. This means that the solenoidal $E^s$ dominates the energy spectrum for the wavenumber within $4\leq k\leq 20$ as stated in Wang \emph{et al.}(2013), that is $E=E^c+E^s=k^{-5/3}(c^ck^{-8/9}+c^s)\approx c^sk^{-5/3}$, and will overall lead the total kinetic energy spectrum $E$ to exhibit of a $-\frac{5}{3}$ scaling as stated by Wang \cite{wang2013}.


\section{Conclusions}

In order to predict scaling laws of compressible turbulence, we applied dimensional analysis to the problem. The following scaling laws were obtained.

(1) The scaling laws of energy spectrum based on velocity $\bm u$ is $$E(k,\varepsilon,M)=\varepsilon^{2/3}k^{-5/3}f[(k\eta)^{4/3},M]$$
which can be estimated as $$E(k,\varepsilon,M)=1.5\nu^{\frac{\ln 6}{12\ln{M}}}\varepsilon^{(\frac{2}{3}-\frac{1}{4}\frac{\ln 6}{\ln{M}})}k^{(-\frac{5}{3}-\frac{1}{3}\frac{\ln 6}{\ln{M}})}$$
(2) The scaling laws of density-weighted energy spectrum $E_\rho(k,\varepsilon,\rho)$ based on density-weighted velocity $\rho^{1/3} \bm u$ is
$$E_\rho(k,\varepsilon,M)=\rho^{1/3}\varepsilon^{2/3}k^{-5/3}F[(k\eta)^{4/3},M]$$
which can be further simplified as $$E_\rho(k,\varepsilon,M)=A_\rho \rho^{1/3}\nu^{-\frac{\beta_\rho}{3\ln{M}}}\varepsilon^{(\frac{2}{3}+\frac{\beta_\rho}{\ln{M}})}k^{(-\frac{5}{3}+\frac{4\beta_\rho}{3\ln{M}})}$$

(3) The intrinsic relationship between the $E(k,\varepsilon,M)$ and $E_\rho(k,\varepsilon,M)$ is $$\frac{E_\rho(k,\varepsilon,M)}{E(k,\varepsilon,M)} = \rho^{1/3}\frac{F(k\eta,M)}{f(k\eta,M)}$$
Theoretically, this paper formulated the scaling laws for compressible turbulence flow by using Barenblatt's incomplete similarity theory. The predicted relations may benefit the understanding of compressible turbulence. It might be worth to mentioning that the scaling laws proposed in this article should be verified again by experiments and more numerical simulations.

The methodology proposed in this paper can be used for other problem, such as, the temporal scaling laws of turbulence \cite{sun2016b}.

\section{Acknowledgement}

Financial supports by the South Africa National Research Foundation (NRF) and travel supports by the State Key Laboratory for Turbulence and Complex Systems at Peking University are gratefully acknowledged. From the bottom of heart, the author would like to thank all reviewers for their constructive criticism and high level academic suggestions, their advice and encouragement inspired me deeply. The author also wish to take this opportunity to appreciate the fruitful discussions on turbulence with Academician U. Frish and S.Y. Chen, Profs. Z.S. She, C.B. Lee and Y.P. Shi.



\begin{thebibliography}{99}

\bibitem{kri}
Kritsuk, A.G., Norman, M.L., Padoan, P. and Wagner, R. The statistics of supersonic isothermal turbulence. \emph{Astrophys.J.} {665}, 416-431(2007).

\bibitem{frisch}
Frisch, U. \textit{Turbulence: The Legacy of A.N. Kolmogorov}. Cambridge University Press,Cambridge(2008).

\bibitem{lee}
Lee, C.B. and Wu, J.Z. Transition in wall-bounded flows. \emph{Appl. Mech. Rev.}, \textbf{61}, 1-20(2008).

\bibitem{she}
She, Z.S., Chen, X., Wu, Y. and Hussain, F. New perspective in statistical modeling of wall-bounded turbulence. \emph{Acta Mech.Sin.}, {26}, :847-861(2010).

\bibitem{zhang1}
Zhang, Y.S., Bi, W.T., Hussain, F., Li, X.L. and She, Z.S. Mach-Number-Invariant Mean-Velocity Profile of Compressible Turbulent Boundary Layers. \emph{Phys. Rev. Lett.}, {109}, 054502(2012).

\bibitem{zhou}
Zhou, H. and Zhang H.X. What is the essence of the so-called century lasting difficult problem in classic physics, the ¡°problem of turbulence"? \emph{Scienta Sinica: Physica, Mechanica and Astronomica}, \textbf{42}, Issue (1): 1-5(2012).

\bibitem{wang2013}
Wang, J., Yang, Y., Shi, Y., Xiao, Z., He, X.T. and Chen, S. Cascade of kinetic energy in three-dimensional compressible turbulence. \emph{Phys. Rev.Lett.} {110}, 214505(2013).

\bibitem{chen2015}
Chen, S.Y., Xia, Z.H., Wang, J.C. and Yang, Y.T., Recent progress in compressible turbulence. \emph{Acta Mech Sin}, 31(3):275-291 (2015).

\bibitem{k41a}
Kolmogorov, A.N. The local structure of turbulence in incompressible viscous fluid for very large Reynolds number. \emph{Dokl. Akad. Nauk SSSR}, {30}, 299-303(1941a) (reprinted in \emph{Proc.R.Soc.Lond.} A, 434,9-13, 1991).

\bibitem{k41b}
Kolmogorov, A.N. On degeneration (decay) of isotropic turbulence in an incompressible visous liquid. \emph{Dokl. Akad. Nauk SSSR}, {31}, 538-540(1941b).

\bibitem{k41c}
Kolmogorov, A.N. Dissipation of energy in locally isotropic turbulence. \emph{Dokl.Akad. Nauk SSSR.}, {32}, 16-18 (1941c).(reprinted in \emph{Proc.R.Soc.Lond.} A, 434,15-17, 1991).

\bibitem{alu2011}
Aluie, H. Compressible turbulence: the cascade and its locality. \emph{Phys.Rev.Lett.}, {106}, 174502(2011).

\bibitem{alu2012}
Aluie, H., Li, S. and Li,H. Conservative cascade of kinetic energy in compressible turbulence. \emph{Astrophys.J.Lett.}, {751}, L29(2012).

\bibitem{alu2013}
Aluie, H. Scale decomposition in compressible turbulence. \emph{Physica D}, {247}, 54-65(2013).

\bibitem{ams}
Armstrong, J.W., Rickett, B.J. and Spangler, S.R. Electron density power spectrum in the local interstellar medium. \emph{Astrophys.J.} {443}, 209-221(1995).

\bibitem{cardy}
Cardy, J., Falkovich, G. and Gawedzki, K. \textit{Nonequilibrium Statistical Mechanics and Turbulence}. Cambridge University Press, Cambridge(2008).

\bibitem{chu}
Chu, B.T. and Kovasznay,L.S.G. Non-linear interactions in a viscous heatconducting compressible gas. \emph{Journal of Fluid Mechanics}, {3}(05):494-514(1958).

\bibitem{fed}
Federrath, C., Roman-Duval, J., Klessen, R.S., Schmidt, W. and Mac Low, M.-M. Comparing the statistics of interstellar turbulence in simulations and observations: Solenoidal versus compressive turbulence forcing. \emph{Astronomy and Astrophysics}, 512, A81(2010).

\bibitem{wang2014}
Wang, J. Cascade of kinetic energy and thermodynamic process in compressible turbulence. \emph{Postdoctoral Research Report}, {Peking University}(2014).

\bibitem{schmidt}
Schmidt, W., Federrath, C. and Klessen, R. Is the scaling of supersonic turbulence universal?  \emph{Phys.Rev.Lett.}, {101}, 194505 (2008).

\bibitem{gal}
Galtier, S. and Banerjee, S. Exact relation for correlation functions in compressible isothermal turbulence. \emph{Phys.Rev.Lett.} {107}, 134501(2011).

\bibitem{sun2015}
Sun, B. The spatial scaling laws of compressible turbulence. arXiv:1502.02815v2 [physics.flu-dyn](2015).

\bibitem{kov}
Kovaznay, L.S.G. Turbulence in supersonic flow. \emph{J. Aero. Sci.}, {20}, 657-682(1953).

\bibitem{light}
Lighthill, M.J. and Whitham, G.B. The effect of compressibility on turbulence, Chapter 22 in Gas Dynamics of Cosmic Clouds. \textit{Proc. 2nd IAU Symposium on Gas Dynamics of Cosmic Clouds, 121-130}, (ed H. G. van Hulst and J. M. Burgess), North Holland Publishing Company, Amsterdam(1955).

\bibitem{moiseev}
Moiseev, S.S., Toor, A.V. and Yanovsky, V.V. The decay of turbulence in the burgers model. \emph{Physica D}, {2}, 187-193(1981).

\bibitem{kad}
Kadomtsev, B.B. and Petviashvili, V.I. Acoustic turbulence. \emph{Soviet Physics Doklady}, {18}, 115-118(1973).

\bibitem{shiva}
Shivamoggi, B.K. Multifractal aspects of the scaling laws in fully developed compressible turbulence. \emph{Annals of Physics}, {243},169-176(1995).


\bibitem{bridgman1922}
Bridgman, P.W. \emph{Dimensional Analysis}. Yale University Press, New Haven(1922).

\bibitem{sedov}
Sedov, L.I. \emph{Similarity and Dimensional Analysis in Mechanics}. Academic Press, New Yrok(1959)

\bibitem{baren}
Barenblatt, G.I. \textit{Similarity, Self-similarity and Intermediate Asymptotics}. Cambridge University Press, Cambridge(1996).

\bibitem{cantwell}
Cantwell, B.J. \emph{Introduction to Symmetry Analysis}. Cambridge University Press, Cambridge(2002).

\bibitem{sun2016}
Sun, B. \emph{Dimensional Analysis and Lie Group} (In Chinese). China High Education Press, Bejing(2016).

\bibitem{liep}
Liepmann, H.W. and Roshko, A. \textit{Elements of Gasdynamics}. Dover Publications, New York(1993).

\bibitem{sun2016b}
Sun, B. The temporal scaling laws of compressible turbulence. \emph{Modern Physics Letters B}. \textbf{30}(23) 1650297 (14 pages)(2016)


\end{thebibliography}
\end{document}